\documentclass[aip,reprint]{revtex4-1}
\usepackage{graphicx}

\newcommand{\kt}{k_\mathrm{B}T}
\renewcommand{\r}{\mathrm}
\renewcommand{\d}{\mathrm{d}}

\newcommand{\eqref}[1]{\textup{(\ref{#1})}}

\begin{document}


\title{Predictions of homogeneous nucleation rates for n-alkanes accounting for the diffuse phase interface and capillary waves}




\author{Barbora Plankov\'{a}}
\email{barbora.plankova@gmail.com}
\affiliation{Institute of Thermomechanics of the CAS, v. v. i., Dolej\v{s}kova 5, 182 00 Prague, Czech Republic}
\affiliation{Department of Mathematics, Faculty of Nuclear Sciences and Physical Engineering, Czech Technical University in Prague, Trojanova 13, 120 00 Prague, Czech Republic}
\author{V\'{a}clav Vin\v{s}}
\author{Jan Hrub\'{y}}
\affiliation{Institute of Thermomechanics of the CAS, v. v. i., Dolej\v{s}kova 5, 182 00 Prague, Czech Republic}


\date{\today}

\begin{abstract}
Homogeneous droplet nucleation has been studied for almost a century but has not yet been fully understood. In this work, we used the density gradient theory (DGT) and considered the influence of capillary waves (CW) on the predicted size-dependent surface tensions and nucleation rates for selected $n$-alkanes. The DGT model was completed by an equation of state (EoS) based on the perturbed-chain statistical associating fluid theory (PC-SAFT) and compared to the classical nucleation theory and the Peng--Robinson EoS. It was found that the critical clusters are practically free of CW because they are so small that even the smallest CW wavelengths do not fit into their finite dimensions. The CW contribute to the entropy of the system and thus decrease the surface tension. A correction for the effect of CW on the surface tension is presented. Effect of the different EoSs is relatively small because by a fortuitous coincidence their predictions are similar in the relevant range of critical cluster sizes. The difference of the DGT predictions to the classical nucleation theory computations is important but not decisive. Of the effects investigated, the most pronounced is the suppression of CW which causes a sizable decrease of the predicted nucleation rates. The major difference between experimental nucleation rate data and theoretical predictions remains in the temperature dependence. For normal alkanes, this discrepancy is much stronger than observed, e.g., for water. Theoretical corrections developed here have a minor influence on the temperature dependency. We provide empirical equations correcting the predicted nucleation rates to values comparable with experiments.
\footnote{This article was published at Journal of Chemical Physics, \url{https://aip.scitation.org/doi/abs/10.1063/1.5008612}. Please cite as B. Plankov\'{a}, V. Vin\v{s}, J. Hrub\'{y}, \textit{J. Chem. Phys.} 147, 164702 (2017)}
\end{abstract}

\pacs{64.60.Q-,64.60.qj,82.60.Nh}

\maketitle 

\section{Introduction} 
Understanding the homogeneous droplet nucleation is important in many natural and industrial processes such as formation of secondary aerosols in the atmosphere, formation of water droplets in the steam turbines, or nucleation during the gas-cleaning procedures, e.g., in processing of natural gas or in technologies for carbon capture and storage. 

Despite many attempts that resolved partial problems of nucleation, there is no complete theory which would give quantitatively correct predictions. 
The basic approach to nucleation is through the classical nucleation theory (CNT), developed by Becker and D\"{o}ring\cite{Becker1935a} and extended by Zeldovich.\cite{Zeldovich1942,Zeldovich1943} The concepts of CNT have been described in several textbooks.\cite{Vehkamaki2006,Kaschiev2000,Kalikmanov2013} The CNT models a cluster as a macroscopic liquid droplet with a step-wise phase interface. However, the real thickness of the interface is comparable with the size of the droplets during the nucleation (in the range of nanometers). The goal of this work is to overcome this limitation by considering a diffuse phase interface (a smooth density profile) and by taking the capillary waves into account. The progress of nucleation theory is backed by experimental results. The state of the art in the experimental nucleation science has been recently reviewed by Wyslouzil and W\"{o}lk.\cite{Wyslouzil2016} In this work, experimental results are used for comparisons without considering experimental circumstances.

Clearly, the step-wise density profile is a crude approximation. A smooth (or diffuse) change of the density between the two phases describes reality better. This approach is considered in the classical density functional theory (DFT) and its approximation, the density gradient theory (DGT). The DGT was first developed in pioneering work of van der Waals,\cite{Waals1893,Rowlinson1979} and later thoroughly elaborated by Cahn and Hilliard.\cite{Cahn1958,Cahn1959} The classical DFT  was proposed by Ebner\cite{Ebner1976} and developed by Evans\cite{Evans1979}, based on the quantum DFT.
%
Laaksonen and Oxtoby\cite{Laaksonen1995} studied nucleation of binary mixtures using Lennard--Jones fluids. They used DFT for ideal and non-ideal mixtures not obeying the Berthelot mixing rules. They found large discrepancies between DFT and CNT, especially for the non-ideal mixtures.
Wilemski and Li\cite{Wilemski2004} investigated DGT and mean field DFT of droplets and bubbles near the spinodal. They compared their results to an approximation of DGT near spinodal developed by Cahn and Hilliard \cite{Cahn1959} and found a qualitative agreement. They address limitations of the mean field treatment of real fluids near spinodal since this region also coincides with region of large fluctuations which are not properly represented in the mean-field theory.
Lutsko\cite{Lutsko2008} examined nucleation using the minimal energy free path method for the Lennard--Jones fluid. He incorporated the nudged-elastic band technique which adds fictitious elastic forces to have collection of density profiles evenly spaced. The advantage of this technique is its relatively simple implementation and robustness. Lutsko got good agreement with simulations of the structure of the cluster, such as the excess number of molecules of the critical cluster or density profiles. More recently,\cite{Lutsko2011} Lutsko incorporated DGT with density profiles approximated as piecewise linear functions for the Lennard--Jones fluid. Unlike the present work, he computed the needed influence parameter using the direct correlation function. Simplified density profiles allowed direct comparison with the CNT. Density profiles, excess number of molecules in the cluster and excess free energy computed by Lutsko agreed well with the simulation data.
McGrath et al.\cite{McGrath2010} examined the nucleation of argon using DFT and Monte Carlo simulations and compared them to CNT. They found rather large differences for four various potentials, as reported by other studies. DFT results agreed more with Monte Carlo simulations than with CNT. They concluded that one of the reasons why CNT fails to describe nucleation correctly is in improper molecular potentials.
Obeidat et al.\cite{Obeidat2010} combined DGT and SAFT (statistical associating fluid theory)\cite{Chapman1989} EoS (equation of state) for methanol and compared it to CNT. To get the results of DGT, they used a finite difference scheme, unlike in this work. As will be described in Sec. \ref{sec_method}, we found this method impractical for the systems considered here. McGrath et al. reported that the DGT gives better nucleation rates than the CNT. While the smooth-interface models are clearly closer to reality than CNT and may provide a significant improvement for specific systems, we found that the considering solely the smooth-interface effect is not sufficient to provide quantitatively correct predictions for $n$-alkanes.


The diffuse-interface approaches of DFT and DGT generate an ``intrinsic'' profile of density through the phase interface. This profile corresponds to averaging over short space of time scales. Besides that, the interface is also disturbed by thermally driven mesoscopic undulations, the capillary waves (CW). The physics of CW was described by Buff, Lovett and Stillinger.\cite{Buff1965}
CW were later investigated by many researchers, e.g., by 
Kayser\cite{Kayser1986} who calculated an increase in interfacial tension caused by cutting off CW longer than a certain wavelength, Beysens and Robert\cite{Beysens1987} and Meunier\cite{Meunier1982,Meunier1987} who combined smoothly changing density approach with CW. Meunier's approach, developed for planar phase interfaces, is used in this work to estimate the effect of CW on the surface tension for droplets (clusters) relevant to nucleation.
Mecke, Dietrich and Napi\'{o}rkowski\cite{Napiorkowski1993,Mecke1999,Mecke2005} derived an effective Hamiltonian combining the effect of DFT and CW and surface tension depending on the wavenumber $q$. Their complex CW model differs significantly from the traditional\cite{Buff1965} approach. Conclusions of their theory led to a rather wide discussion.\cite{Doerr1999,Fradin2000,Lin2005,Tarazona2007,Hiester2008}
Chac\'{o}n, Tarazona and Alejandre\cite{Chacon2003,Tarazona2004,Chacon2006} developed a method to determine a definition of the instantaneous dividing surface in molecular simulations applied to atomic fluids and water. They bridged the approaches of CW and interpretation of x-ray reflectivity experiments making it an alternative model of CW. 
Boltachev and Baidakov\cite{Boltachev2006} provided a new empirical model of CW fitted to  ellipsometry data for argon. Their approach provides an alternative for determining the effect of CW on surface tension and thickness of the phase interface. 

In this work we address the problem of homogeneous droplet nucleation with DGT using PC (perturbed chain) SAFT\cite{Gross2000,Gross2001} EoS and considering effect of CW. For comparison, we add theoretical results also for CNT and cubic Peng--Robinson (PR)\cite{Peng1976} EoS.

This article is organized as follows. In Sec.~\ref{sec_nucleation} we consider some basic relations for nucleation computations and mention their weak points. We connect the nucleation rate to the work needed to form the droplet in supersaturated vapor. In Sec.~\ref{sec_dgt} we provide mathematical results of DGT to determine the density profiles that correspond to the work of formation of a cluster. In Sec.~\ref{sec_pc_saft} we introduce the PC-SAFT EoS. In Sec. \ref{sec_cw} we characterize the impact of CW on surface tension and show how to eliminate CW from the input of DGT. In Sec.~\ref{sec_method} we explain mathematical methods for the numerical solution defined in Sec. \ref{sec_dgt}. In Sec.~\ref{sec_results} we provide the resulting nucleation rates of four $n$-alkanes; namely $n$-heptane, $n$-octane, $n$-nonane and $n$-decane, and compare them to the experimental data. We reevaluate the data to determine the temperature dependence of the supersaturation and number of molecules in the critical cluster. We examine impact of CW on the surface tension and the nucleation rate. In the last section, Sec. \ref{sec_conclusions}, we draw conclusions about the nucleation rates, temperature dependence, and influence of CW.

\section{Nucleation rates}
\label{sec_nucleation}
Nucleation of droplets in a homogeneous vapor can occur if the vapor density $\rho_\mathrm{V}$ is greater than the density of saturated vapor $\rho_{\mathrm{V\infty}}$ at a given temperature $T$. In further text, subscript $\mathrm{\infty}$ refers to the infinite curvature radius of the planar phase interface under the saturation conditions. The departure of the system from saturation is expressed in terms of supersaturation $S$ defined as
\begin{equation}
 S = \exp\left(\frac{\mu_\mathrm{V} - \mu_\infty}{k_\mathrm{B}T}\right),
\label{eq:S}
\end{equation}
where $\mu_\mathrm{V}$ is the chemical potential of the vapor phase, $\mu_\infty$ is the chemical potential at saturation, which is equal for liquid and vapor phases, and $k_\mathrm{B}$ is the Boltzmann constant. 
If $S>1$, the vapor becomes supersaturated. Supersaturated vapor is in a metastable state, i.e., it is stable with respect to small fluctuations but unstable with respect to large fluctuations, particularly in the form of clusters of the newly forming phase.\cite{Rowlinson1982,Kalikmanov2013} Homogeneous droplet nucleation is the process of formation of liquid droplets in the supersaturated vapor phase free of aerosol particles, impurities, or solid walls.
These droplets can be viewed as clusters of $n$ molecules, referred to as $n$-mers. Microscopically, these clusters grow or shrink if a monomer joins or leaves the cluster. Condensation and evaporation of dimers is less probable except for strongly associating vapors. 
Of all the $n$-mers, the most important for nucleation is the so-called critical cluster with $n_\r c$ molecules which has the same probability to grow as to shrink.\cite{Becker1935a} In terms of thermodynamics, this condition can be interpreted as an equality of chemical potential in the liquid core of the critical cluster and the chemical potential of the supersaturated vapor,
\begin{equation}\label{eq_mu}
    \mu_{\r L}(n_\r c) = \mu_\r V \,.
\end{equation}
The number of droplets formed in a unit of volume per a unit of time is called the nucleation rate, $J$. Considering the chain of individual growth-shrink processes, it is possible to derive an expression for the nucleation rate in the form of a converging infinite sum,\cite{Vehkamaki2006,Kaschiev2000,Kalikmanov2013}
\begin{equation}
 J = \left[ \sum_{n=1}^{\infty} \frac{1}{f_nC_n^\mathrm{eq}} \right]^{-1},
\label{eq_J_sum}
\end{equation}
where $f_n$ is the condensation rate, i.e., the number of monomers impacting on the surface of $n$-mer per unit of time; the sticking probability is taken as unity. $C_n^\mathrm{eq}$ is the equilibrium concentration of $n$-mers at given $T$ and $\mu_\r V$ in a system which is constrained in the way that the growth over certain size $n_\r {max}$ is prohibited. If $n_\r {max}$ is substantially larger than the critical size $n_\r c$, its magnitude does not play any role. When no interactions between large clusters are considered, the limit of maximal size of $n$-mer can be placed to infinity,\cite{Vehkamaki2006} as adopted in Eq. \eqref{eq_J_sum}. Constrained equilibrium concentrations enter the nucleation theory via the condition of detailed balance\cite{Vehkamaki2006,Kalikmanov2013} stating that in the equilibrium all microscopic processes must have their time-mirrored counterparts which are equally probable. This condition allows for expression of the evaporation rate in terms of the condensation rate. The latter is given as
\begin{equation}
 f_{n} = \nu A_n,
\end{equation}
where $A_n$ is the surface area of $n$-mer and $\nu$ is the impingement rate of monomers per surface area. This is given by the kinetic theory of gases as\cite{Vehkamaki2006,Kalikmanov2013,Landau1980}
\begin{equation}\label{eq_nu}
 \nu = C_1\sqrt{\frac{\kt}{2\pi M}},
\end{equation}
where $C_1$ is the concentration of monomers and $M$ is the molar mass.
The concentration $C_n^\mathrm{eq}$ can be expressed in terms of the work of formation $\Delta \Omega_n$ of $n$-mer as\cite{Vehkamaki2006,Kalikmanov2013}
\begin{equation}
 C_n^\mathrm{eq} = \frac1\vartheta \exp\Big( -\frac{\Delta\Omega_n}{\kt} \Big), 
\label{eq_C_n_eq}
\end{equation}
where $\vartheta$ is the volume available for the translation of the cluster and attributes to the translational degrees of freedom of the cluster. The Boltzmann factor $\exp(-\Delta \Omega_n/\kt)$ can be interpreted as a probability of observing at least one $n$-mer in a region of volume $\vartheta$ in a (constrained) equilibrium system. The original choice by Becker and D\"oring\cite{Becker1935a} was $1/\vartheta=\rho_{\r V}$, which, however, has been shown to affect the dependence of cluster concentrations on supersaturation in a theoretically unacceptable manner.\cite{Barrett1997,Girshick1990,Hruby2004,Reguera2004} In this work, we adopt $1/\vartheta=\rho_{\r V\infty}$, corresponding to the correction by Courtney.\cite{Courtney1961} The CNT modification by Girshick and Chiu\cite{Girshick1990} corresponds to $1/\vartheta=\rho_{\r V\infty}e^\theta$, where $\theta$ is the so-called dimensionless surface tension (variable consisting of the surface tension, liquid density, and thermal energy $\kt$). The reasoning behind the Girshick-Chiu modification is that the expression for equilibrium concentration of clusters, Eq.~(\ref{eq_C_n_eq}), should hold down to monomers -- provided that the dimensionless surface tension is size-independent. Here, however, we consider the size dependency and, therefore, this correction was not adopted.

The sum in Eq. \eqref{eq_J_sum} can be approximated by an integral. The integrand is a function of $n$ showing a sharp peak at $n_\r c$, which is due to the exponential part (the work of formation) in Eq.~\eqref{eq_C_n_eq}.
Compared to that, the condensation rate varies modestly with $n$ and, therefore, can be approximated as $f_{n}\simeq f_{n_\r c}$.\cite{Vehkamaki2006,Kaschiev2000,Kalikmanov2013} In a more refined treatment, $1/f_n$ can be approximated as a linear function of $n$ in the neighborhood of $n_\r c$. The proportional part cancels out upon the integration. The peaked function $1/C_n^\mathrm{eq}$ is then replaced by a Gaussian function and, integrated, yielding to\cite{Vehkamaki2006,Kaschiev2000,Kalikmanov2013}
\begin{equation}\label{eq_J_gen}
 J = \frac1\vartheta A(n_\r c)\nu \mathcal{Z}  \exp\left(-\frac{\Delta \Omega(n_\r c)}{\kt}\right),
\end{equation}
where $\mathcal{Z}$ is the Zeldovich factor defined as
\begin{equation}
 \mathcal{Z} = \sqrt{-\frac{\Delta\Omega''(n_\r c)}{2\pi\kt}}
\simeq \frac{2}{\rho_\r{L,\infty}} \sqrt{\frac{\sigma_\infty}{\kt}}.
\label{eq_Zeldovich}
\end{equation}
In Eq. \eqref{eq_Zeldovich}, $\Delta\Omega''$ is the second derivative of the work of formation with respect to the number of molecules in the cluster, evaluated for the critical cluster, $\Delta\Omega''(n_\r c)=\partial^2\Delta\Omega/\partial n^2 (n_\r c)$, $\rho_\r{L,\infty}$ is the liquid density, and $\sigma_\infty$ is the surface tension for the saturated state. The rightmost expression of the Zeldovich factor is an approximation based on CNT. Although different expressions will be used for $\Delta\Omega$, the Zeldovich factor can be kept in the classical form because it is rather insensitive to the modifications considered. With Eqs.~(\ref{eq_nu}) and (\ref{eq_Zeldovich}), and using $\vartheta$ corresponding to Courtney's correction,\cite{Courtney1961} nucleation rate given by Eq. (\ref{eq_J_gen}) becomes
\begin{equation}
 J = \frac{\rho_\mathrm{V}\rho_{V\infty}}{\rho_{\mathrm{L \infty}}} \sqrt{\frac{2\sigma_{\infty}}{\pi M}} \exp\left(-\frac{\Delta\Omega}{k_\mathrm{B}T} \right)\,. \label{eq_J_ic}
\end{equation}


The work of formation $\Delta \Omega$ of the critical cluster can be expressed in terms of the Gibbsian thermodynamics of phase interfaces as
\begin{equation}
  \Delta \Omega = \frac 13 A_\mathrm{s} \sigma,
\label{eq_Delta_Omega2}
\end{equation}
where $A_\mathrm{s} = 4\pi r_\mathrm{s}^2$ is the area of the surface of tension of the cluster and $r_\r s$ is its radius. The surface of tension is defined by postulating the Young--Laplace equation in its standard form,\cite{Rowlinson1982,Vehkamaki2006}
\begin{equation}
 \Delta p = \frac{2\sigma}{r_\r s},
\label{eq_young}
\end{equation}
where $\Delta p = p_\r L - p_\r V$ is the Laplace pressure.
The work of formation is connected to the surface tension via
\begin{equation}\label{eq:DOcnt}
 \Delta \Omega = \frac{16\pi}{3} \frac{\sigma^3}{\Delta p^2}.
\end{equation}
We note that the Laplace pressure $\Delta p$ for the critical cluster is related to supersaturation given by Eq. (\ref{eq:S}) through the equality defined in Eq. (\ref{eq_mu}) and EoS employed for calculation of chemical potential and density.

The work of formation for the CNT is obtained by substituting the flat-interface surface tension $\sigma_\infty$ into Eq.~(\ref{eq:DOcnt}). Inversely, if $\Delta \Omega$ is known, this equation can be used to compute the (size-dependent) surface tension for the critical cluster. Typically, further simplifications are employed in CNT. In particular, the compressibility of the liquid is neglected and the gas phase is assumed as an ideal gas. These simplifications were not employed in this work; the CNT work of formation was computed from appropriate EoS (PR EoS and PC-SAFT EoS).

\section{Density gradient theory (DGT)}
\label{sec_dgt}
The DGT operates with a density $\rho$ smoothly changing throughout the phase interface. The local Helmholtz energy density is assumed\cite{Cahn1958} as a function of density and invariants of the spatial derivatives of density -- the gradient squared $(\nabla \rho)^2$, and the Laplacian $\nabla^2 \rho$.
The Helmholtz energy is expanded in terms of these invariants around the Helmholtz energy of the homogeneous fluid $f^0(\rho)$ which can be computed from EoS in the region between saturated vapor and saturated liquid. We note that this region involves metastable vapor, metastable liquid, and the region between spinodals, which is unstable in the homogeneous case, but inside the phase interface it is stabilized by the density gradient.
Retaining only linear terms in a Taylor expansion with respect to variables $(\nabla \rho)^2$ and $\nabla^2 \rho$, the Helmholtz energy density reads\cite{Cahn1958,Cahn1959,Baidakov2002} 
\begin{equation}
 f(\rho,\nabla\rho,\nabla^2 \rho) = f^0(\rho) + c_1(\rho,T) \nabla^2 \rho + c_2(\rho,T) (\nabla \rho)^2 + \ldots,
\label{eq_dft_dgt_f}
\end{equation}
where $c_1$ and $c_2$ are Taylor coefficients. Using the Green (divergence) theorem, the second derivative $\nabla^2$ in Eq. \eqref{eq_dft_dgt_f} can be substituted to the first derivative which leads to\cite{Cahn1958}
\begin{equation}
 f(\rho) = f^0(\rho) + c(\rho,T) \cdot (\nabla \rho)^2 + \ldots .
\label{eq_f1}
\end{equation}
In Eq. \eqref{eq_f1}, the second term accounts for the inhomogeneity induced by the interface. Coefficient $c= -\partial c_1/\partial\rho + c_2$ is called the influence parameter\cite{Cahn1958} and can be related to the direct correlation function.\cite{Davis1996} In this work, it will be evaluated from the experimental data (as discussed below) and assumed as density-independent.

The work of formation of a droplet according to DGT can be expressed as a volume integral consisting of homogeneous and gradient parts,\cite{Cahn1958,Davis1996}
\begin{equation}
 \Delta \Omega [\rho(r)] = \int_0^{\infty} \left[ \Delta \omega^0(\rho) + \frac12c\left( \frac{\d\rho}{\d r}\right)^2 \right] 4\pi r^2 \d r,
\label{eq_Delta_Omega1}
\end{equation}
where $r$ is the radial coordinate (the distance from the center of the droplet). Term $\Delta\omega^0$ is the difference of the grand-potential density of the inhomogeneous system with respect to the homogeneous one at the same chemical potential (equal to the negative value of the vapor pressure $p_\r V$),
\begin{equation}
 \Delta\omega^0(\rho) = f^0(\rho) - \rho \mu_\mathrm{V} + p_\r V.
\end{equation}
We note that the work of formation, Eq. \eqref{eq_Delta_Omega1}, is not a mere function of the density, but a \textit{functional} since its value depends on the whole density profile $\rho(r)$ and not just on the local value.
The work of formation has a saddle point for the density profile corresponding to the so-called critical cluster; the work of formation of the critical cluster is minimal with respect to all the properties in the functional space except for density profile variations changing the size of the cluster (shifting the phase interface closer to or further from the center), for which it is maximal.\cite{Rowlinson1982,Barrett1997} Setting the functional derivative of
Eq.~\eqref{eq_Delta_Omega1} equal to zero as a condition for the saddle point leads to an Euler--Lagrange equation\cite{Cahn1959}
\begin{equation}
  \frac{\d^2\rho}{\d r^2} + \frac2r \frac{\d\rho}{\d r} = \frac{1}{c} \Delta \mu(\rho).
\label{eq_diff_eq}
\end{equation}
In Eq. \eqref{eq_diff_eq}, $\Delta\mu = \mu^0(\rho) - \mu_\mathrm{V}$ is the difference of chemical potential of a hypothetical homogeneous fluid at a given local density $\rho$ and the actual chemical potential $\mu_\mathrm{V}$. In the critical cluster, the chemical potential is everywhere equal to chemical potential of the homogeneous vapor phase. 
To solve the Euler--Lagrange equation, two boundary conditions are needed,\cite{Cahn1959}
\begin{equation}
  \rho(r\rightarrow \infty) = \rho_\mathrm{V}, \quad \frac{\d\rho}{\d r}(0) = 0.
\label{eq_conditions}
\end{equation}

Other important outputs of the DGT calculations are the excess number of molecules $\Delta N$ and the surface tension of the saturated state $\sigma_\infty$ (corresponding to the planar phase interface). 
The excess number of molecules is a difference of the number of molecules in a system containing a cluster to the number of molecules in a system of the same volume containing homogeneous vapor only. It can be computed as
\begin{equation}
 \Delta N = \int_0^\infty \left[\rho(r)-\rho_\mathrm{V}\right]4\pi r^2 \d r.
 \label{eq_deltaN}
\end{equation}
The surface tension for a planar phase interface separating saturated liquid and saturated vapor can be obtained in a relation similar to Eq. \eqref{eq_Delta_Omega1},
\begin{equation}
  \sigma_{\infty} = \int_{-\infty}^{\infty} \left[ \Delta \omega^0(\rho) + \frac12c\left( \frac{\d\rho}{\d z}\right)^2 \right] \d z,
\label{eq_sigma_inf}
\end{equation}
where $z$ is the coordinate perpendicular to the phase interface. Surface tension can also be obtained from experimental data; therefore, Eq. \eqref{eq_sigma_inf} can be used for computation of the (density-independent) influence parameter $c$.\cite{Hruby2007}


\section{PC-SAFT equation of state}
\label{sec_pc_saft}
In this work, DGT was combined with the PC-SAFT\cite{Gross2000,Gross2001} EoS, which belongs to the family of modern SAFT-type equations of state being developed since 1990's.\cite{Chapman1989,Muller2001,Tan2008}
The SAFT-type equations produce a qualitatively correct isotherm in the two-phase region, the so-called van der Waals loop. On the other hand, they provide remarkably good results for the liquid phase compared, e.g., to the cubic equations. As a consequence, the SAFT-type equations represent a convenient tool for modeling of vapor-liquid phase interfaces and for modeling nucleation of droplets and bubbles using DFT and DGT.

The SAFT-type EoSs are defined in the form of the Helmholtz energy, which is given as a sum of the ideal gas part $F_\r{id}$ evaluated from the isobaric heat capacity of the ideal gas and the residual part $F_\r{res}$ defined by the SAFT terms. An important advantage of the SAFT-type EoSs is that the residual part of the Helmholtz energy is expressed as a sum of individual contributions accounting for various types of intermolecular interactions,\cite{Tan2008} e.g., the van der Waals attractions, Coulombic forces, or hydrogen bonds.\cite{Muller2001,Tan2008} In PC-SAFT, the residual part consists of the hard chain contribution $F_\r{hc}$ representing the reference fluid and the perturbation contribution $F_\r{disp}$.
The hard chain contribution and the perturbation contribution are defined by a set of three molecular parameters: the segment number $m$, the segment diameter $\sigma$, and the energy parameter $\epsilon/k_\r B$. Values of these parameters are usually correlated to the saturation pressure and the liquid density.\cite{Gross2001}
Due to their reasonable physical basis and possibility of considering various contributions in $F_\r{res}$, the SAFT-type equations can be applied to a large variety of substances ranging from relatively simple components such as gases, $n$-alkanes or halogenated hydrocarbons\cite{Vins2009,Vins2013} to polar substances,\cite{Vins2012,Schafer2014} polymers,\cite{Gross2003} and associating fluids.\cite{Tan2008,Grenner2007} Consequently, these equations of state are becoming popular both in scientific and engineering applications.\cite{Muller2001,Tan2008} Molecular parameters $m$, $\sigma$ and $\epsilon/k_\r B$ for selected $n$-alkanes used in this study were taken from the original work of Gross and Sadowski.\cite{Gross2001}

\section{Capillary waves}
\label{sec_cw}
Thermal motion of molecules causes that the interface is not a smooth surface but it is rather disturbed by the capillary waves.\cite{Buff1965} We will use some results of the so-called mode-coupling theory developed by Meunier\cite{Meunier1987} which considers effect of the broadening of the phase interface due to the CW on the surface tension for a planar phase interface.

We define a ``bare'' surface tension $\sigma_\r{bare}$ as the surface tension of a phase interface which is forced to be perfectly planar, cleared of CW. The bare surface tension still accounts for some thermal motion of molecules, which, however, is only correlated over a lengthscale comparable to the bulk correlation length. The surface tension determined in macroscopic experiments $\sigma_{\mathrm{exp}}$ includes the effect of CW. Allowing CW adds some disorder to the system which leads to an increase of the surface excess entropy. Consequently, multiplied by temperature, the CW reduce the surface tension as\cite{Meunier1987}
\begin{equation}
 \sigma_{\mathrm{exp}} = \sigma_\r{bare} - \frac{3}{8\pi} \kt q_\r{max}^2,
\label{eq_sigma_cw1}
\end{equation}
where $q_\r{max}$ is the upper cutoff of the wavenumbers considered. $q_\r{max}$ corresponds to CW with the smallest wavelength that are not already accounted for in DGT.

Meunier\cite{Meunier1987} estimated $q_\r{max}$ based on the mode-coupling theory as
\begin{equation}
 q_\r{max} = \frac{1}{2.64\xi^+},
\label{eq_qmax_xi+}
\end{equation}
where we corrected the original value \cite{Meunier1987} 2.55 to the new value $2.64$ according to the recent results of the critical scaling theory.\cite{Anisimov2010} $\xi^+$ is the correlation length of the supercritical fluid. Since we are dealing with subcritical temperatures, it is more consistent (but numerically equivalent) to reformulate Eq. \eqref{eq_qmax_xi+} in terms of $\xi^-$, the correlation length of the subcritical fluid,
\begin{equation}
 q_\r{max} = \frac{1}{5.17\xi^-}.
\label{eq_qmax}
\end{equation}
Both correlation lengths obey a scaling law
\begin{equation}
 \xi^{\pm} = \xi^{\pm}_0 \left|1-T/T_\r c \right|^{-\nu},
\label{eq_xi}
\end{equation}
where $\nu=0.63$ is a critical exponent. Critical amplitudes $\xi^\pm_0$ in Eq. \eqref{eq_xi} are related as
\begin{equation}
 \frac{\xi_0^+}{\xi_0^-}\simeq1.96,
\label{eq_xi_amplitude_ratio}
\end{equation}
which explains the mathematical equivalence of Eqs.~\eqref{eq_qmax_xi+} and \eqref{eq_qmax}. Critical amplitude $\xi^+_0$ can be obtained from another critical scaling ratio
\begin{equation}
 \frac{\sigma_{0} (\xi^+_0)^2}{\kt_\r c} \simeq 0.39,
\label{eq_sig_amplitude_ratio}
\end{equation}
where $T_\r c$ is the critical temperature. The critical amplitude $\sigma_{0}$ for the surface tension has been obtained by fitting scaling relation
\begin{equation}
    \sigma = \sigma_0 (1-T/T_\r c)^{2\nu}
    \label{eq_sig_scaling}
\end{equation}
to the experimental surface tension data (for the list of data see Appendix \ref{app_sigma}).

Using Eqs. \eqref{eq_qmax} to \eqref{eq_sig_scaling}, the bare surface tension in Eq. \eqref{eq_sigma_cw1} can be related to the experimental value as
\begin{equation}
 \sigma_\r{bare} = \sigma_{\mathrm{exp}} \left(1+ \frac{3}{8\pi}\frac{T}{T_\mathrm{c}}\frac{1}{2.72}\right).
\label{eq_sigma_bare}
\end{equation}
This is equivalent to the result of Gross,\cite{Gross2009} aside from the updated numerical value in the denominator. Effect of the CW is strongest at the critical point, where the bare surface tension is higher by $4.42\,\%$ than the experimental value, and it remains significant down to the triple point temperature. We note that Eq. \eqref{eq_sigma_bare} is universal, i.e., substance independent.

Eq. \eqref{eq_sigma_cw1} assumes an infinitely large system. In a finite system, the maximal wavelength of the CW is limited by the system's linear dimension. Correspondingly, a lower cutoff of the wavenumber $q_\r{min}$ should be considered. This is also the case of the nanodroplets relevant to nucleation. We estimate the maximal wavelength as a half of the circumference of the droplet,
\begin{equation}
 \lambda_\r{max} = \pi r_\r s,
\end{equation}
where radius of the droplet is represented by the radius of the surface of tension, $r_\r s$. This estimate corresponds to the oblate-prolate ellipsoid oscillations, which is the lowest mode of CW on a sphere. The lower cutoff of the wavenumber is then
\begin{equation}
 q_\r{min} = \frac{2\pi}{\lambda_\r{max}} = \frac{2}{r_\r s}.
\label{eq_qmin}
\end{equation}

The critical cluster radius depends on the temperature $T$ and the supersaturation $S$.
We evaluated both cutoffs for the experimental nucleation data considered here for the comparison with theoretical predictions (more in Sec. \ref{sec_results}). This was done in order to determine the range of the CW wavenumbers relevant for the critical clusters.
\begin{figure}
 \centering
 \includegraphics[scale=1]{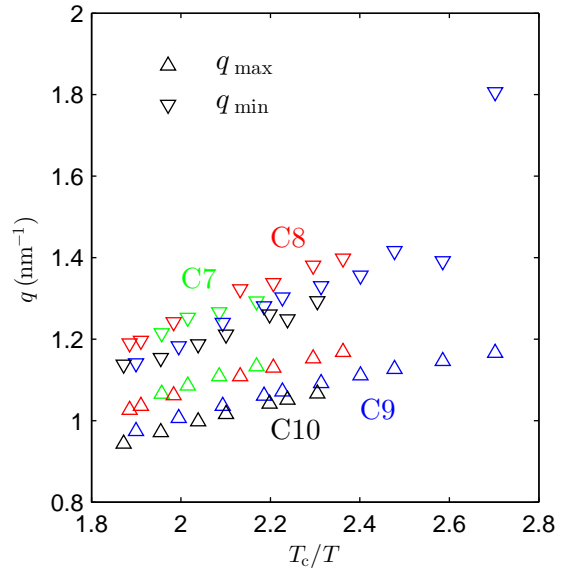}
 \caption{Minimal cutoff $q_\r{min}$, Eq.~\eqref{eq_qmin}, and maximal cutoff $q_\r{max}$, Eq.~\eqref{eq_qmax}, of the CW wave numbers as functions of the inverse reduced temperature. Plotted for $n$-heptane (C7), $n$-octane (C8), $n$-nonane (C9) and $n$-decane (C10). The plotted values correspond to experimental nucleation data.
  }
\label{fig_qmin_qmax}
\end{figure}
The upper cutoff $q_\r{max}$ was directly computed for a given temperature $T$ rounded to a certain value for each series (every individual experiment differed in the temperature somewhat). To evaluate the lower cutoff $q_\r{min}$ using Eq.~\eqref{eq_qmin}, the radius $r_\r s$ was obtained by interpolating the data computed by DGT for the experimental supersaturation $S_\r{exp}$. For every rounded temperature we got several cutoffs $q_\r{min}$ depending on the supersaturation of the data point; for clarity we considered only the minimal $q_\r{min}$ for given rounded temperature so that the range $(q_\r{min},q_\r{max})$ was the widest possible.
Fig.~\ref{fig_qmin_qmax} shows both cutoffs for the experimental data computed using Eqs. \eqref{eq_qmax} and \eqref{eq_qmin}. The seemingly outlaying $q_\r{min}$ point in the upper right corner of Fig. \ref{fig_qmin_qmax} corresponds to data by Wagner and Strey\cite{Wagner1984} at higher supersaturations and thus smaller critical clusters than the other data. It can be seen that $q_\r{min}>q_\r{max}$ in all cases. Therefore, the range $(q_\r{min},q_\r{max})$ is an empty interval; this means that the critical clusters are too small to accommodate any CW. All thermal fluctuations are bulk-like density fluctuations included in the continuous density profile model. Consequently, the bare spherical phase interface obtained by DFT or DGT is an appropriate representation for the critical clusters.

\section{Numerical method}
\label{sec_method}
Eq. \eqref{eq_diff_eq} with boundary conditions given by Eq. \eqref{eq_conditions} form a boundary value problem. This simply looking problem has several difficulties. First, the density profile near the gaseous phase has a sharp, corner-like, shape; at the gas-phase side, its slope changes abruptly. This ``stiff'' behavior is caused by the strongly non-linear right-hand side of the differential equation, Eq. \eqref{eq_diff_eq}. The second difficulty is that for large droplets the density profile in the interior of the droplet is almost perfectly flat. The difference of the density in the droplet center $\rho(0)$ to the density $\rho_\r L$, following from the equality of chemical potentials, Eq. (\ref{eq_mu}), approaches zero rapidly as the droplet becomes large compared to the thickness of the phase interface. This situation occurs for droplets much larger than the typical critical clusters, so that the resulting uncertainty does not influence the nucleation rate computations presented here. However, it represents an obstacle in studying the asymptotic behavior of the size-dependent surface tension when the droplet radius approaches infinity.

In a previous work,\cite{Hruby2007} the boundary value problem was solved using a \textsc{matlab}\cite{Mathworks.com} procedure \textsc{bvp4}. This procedure uses a finite difference scheme: the original interval is divided into subintervals, where the right-hand side of differential equation (\ref{eq_diff_eq}) is integrated. An initial approximation of the solution is required. In many general cases, the initial guess can be very crude. However, in case of Eq. (\ref{eq_diff_eq}) with right-hand side computed from a realistic EoS, this initial approximation had to be very close to the final solution to ensure a convergence of the iteration procedure. Due to these problems, this integration procedure was considered as impractical for routine calculations.

To overcome the above-mentioned difficulties, a new algorithm was developed based on the simple but robust shooting method. The boundary value problem was converted into an initial value problem by estimating the density in the center of the droplet $\rho_0$, so the initial conditions were
\begin{equation}
 \rho(0) = \rho_0, \quad \frac{\d\rho}{\d r}(0) = 0.
\label{eq_conditions2}
\end{equation}
The initial value problem was then solved using an implementation of the variable-step Runge--Kutta method (\textsc{matlab} function \textsc{ode45}).
The original boundary condition in Eq. (\ref{eq_conditions}), in theoretical formulation placed to infinity, can be modified to $\rho(R) = \rho_\mathrm{V}$ where $R$ is finite but sufficiently large, such that the theoretical density profile at $R$ is sufficiently close to the vapor density.
In the shooting method, the center density $\rho(0)$ is searched for in an iteration procedure such that the right-hand boundary condition is matched.

\begin{figure}
 \centering
 \includegraphics[scale=1]{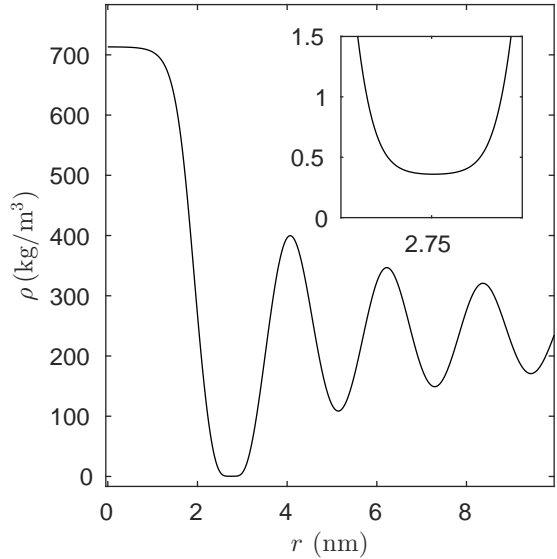}
 \caption{Density profile $\rho$ of a critical cluster as a function of distance from the center of the droplet $r$ for $n$-heptane at temperature $T = 276 \,\r K$ and supersaturation $S = 4.4$ computed using DGT combined with PC-SAFT EoS. Inset is the detail of the minimum around 2.7 nm.
 }
\label{fig_dens_profile}
\end{figure}

As shown in Fig. \ref{fig_dens_profile}, numerical solutions of Eq.~\eqref{eq_diff_eq} with conditions defined by Eq.  \eqref{eq_conditions2} indicate an oscillatory behavior. We note that these oscillations are not a result of a particular numerical procedure but rather they are an intrinsic feature of Euler-Lagrage equation, Eq.~(\ref{eq_diff_eq}). In this example, a decrease from density in the center $\rho_0 = 713 \,\r{kg\,m^{-3}}$ down to the vapor density $\rho_\r V = 0.36 \,\r{kg\,m^{-3}}$ occurs within the first 3~nm. Then the curve detaches from the vapor density and oscillates around density $\rho \simeq 250 \,\r{kg\,m^{-3}}$, corresponding to a physically unstable (but numerically stable) solution of the equality of chemical potentials, Eq.~(\ref{eq_mu}). This oscillatory behavior was observed for estimates of $\rho_0$ lying between the correct value and the saturated liquid density $\rho_\r L$. Solutions of Eq.~(\ref{eq_diff_eq}) for guesses of $\rho_0$ slightly below the correct value approached the vapor density and then detached downwards, quickly crossing the zero density. Because of this behavior, enforcing the boundary condition at certain fixed coordinate $R$ was found impractical. Therefore, the second condition of Eq. \eqref{eq_conditions2} was changed to
\begin{equation}\label{eq:RightCondMod}
 \min_{r<R}\rho(r) = \rho_\r V.
\end{equation}
In addition, estimates undershooting the vapor density have been rejected. As a result, the proper value $\rho_0$ was approached from below (since lower $\rho_0$ corresponded to higher $\rho_\r V$). To achieve a fast convergence, we attempted to compute a next approximation of $\rho_0$ based on the linear extrapolation of the previous two guesses and the corresponding difference of the left and right sides of Eq.~(\ref{eq:RightCondMod}). Unfortunately, this modified secant method appeared to be unreliable because the numerical solution of Eq.~(\ref{eq_diff_eq}) exhibits a quasi-random error when considered as a function of parameter $\rho_0$. The magnitude of this numerical error is small enough so that it does not affect the computations of the work of formation. However, the non-smooth dependence on $\rho_0$ caused convergence failures for $\rho_0$ guesses close to the correct boundary value solution. This problem was remedied by changing the secant method to a slower but more robust bisection method. Additional improvements\cite{Plankova2011} were developed to make the algorithm more robust and precise. In particular, the inner and outer parts of the density profile in intervals, where the compressibility only negligibly differs from compressibility at $\rho_0$ or at $\rho_V$, were substituted by an analytical solution of a linearized Eq. \eqref{eq_diff_eq}.

\section{Results and discussion}
\label{sec_results}
\begin{figure}
 \centering
 \includegraphics[scale=1]{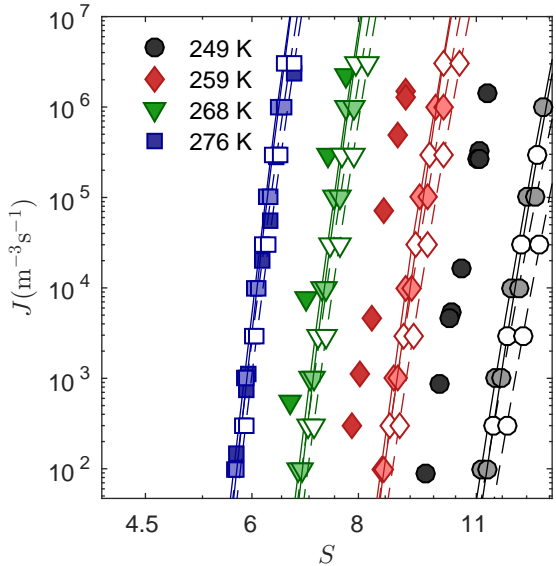}
 \caption{Nucleation rates $J$ of $n$-heptane as functions of supersaturation $S$ for various temperatures in logarithmic scale. One color always corresponds to one temperature. Comparison of theoretical nucleation rates, Eq. \eqref{eq_J_ic}, with experiments by Rudek et\,al.\cite{Rudek1996} Solid lines correspond to DGT, dashed lines to CNT. Lines with light-color symbols corresponds to PC-SAFT EoS, with white symbols to PR EoS. Experimental data are depicted by dark-colored symbols.}
\label{fig1}
\end{figure}
\begin{figure}
 \centering
 \includegraphics[scale=1]{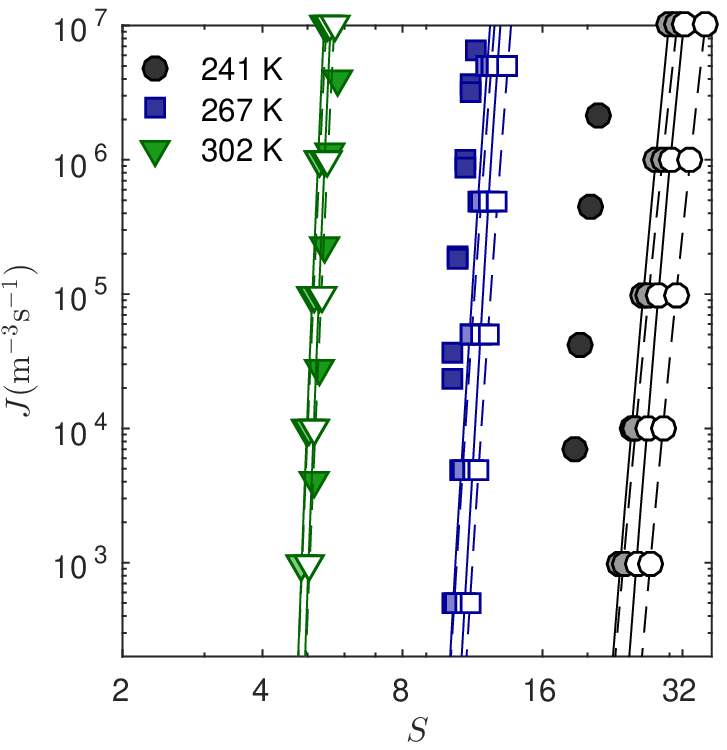}
 \caption{Nucleation rates $J$ of $n$-octane as functions of supersaturation $S$ for various temperatures in logarithmic scale. Experimental data are taken from Rudek et\,al.\cite{Rudek1996} Markers and lines are defined in the same way as in Fig. \ref{fig1}.}
\label{fig2}
\end{figure}
\begin{figure}
 \centering
 \includegraphics[scale=1]{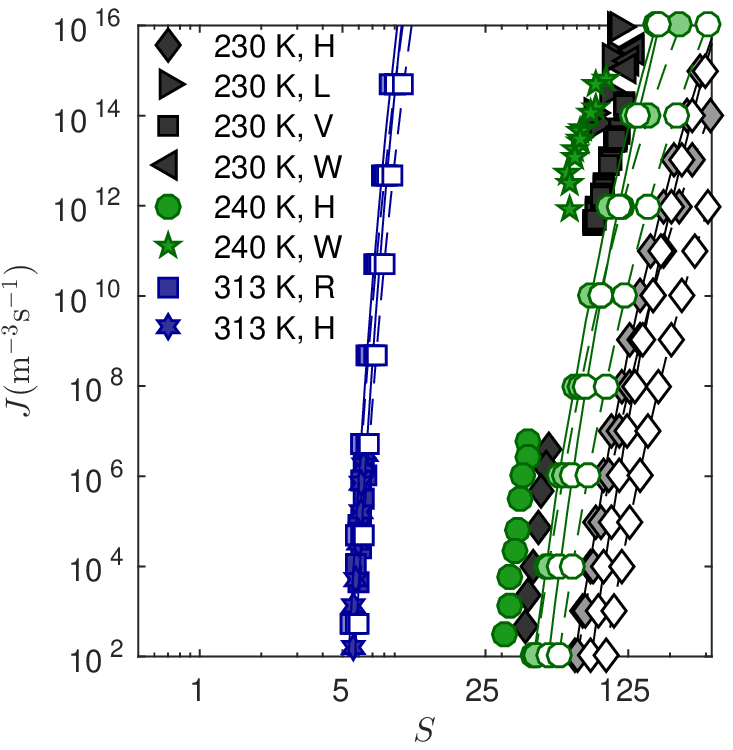}
 \caption{Nucleation rates $J$ of $n$-nonane as functions of supersaturation $S$ for various temperatures in logarithmic scale. Experimental data are taken from Wagner et al.\cite{Wagner1984} (W), Hung et al.\cite{Hung1989} (H), Rudek et al.\cite{Rudek1996} (R), Luijten\cite{Luijten1998} (L) and Viisanen et al.\cite{Viisanen1998} (V). Markers and lines are defined in the same way as in Fig. \ref{fig1}.}
\label{fig3}
\end{figure}
\begin{figure}
 \centering
 \includegraphics[scale=1]{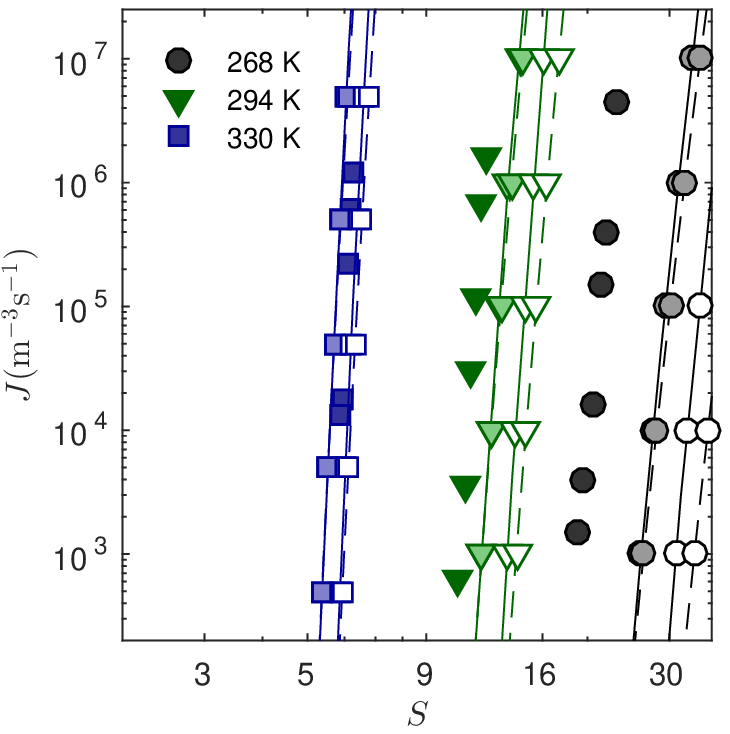}
 \caption{Nucleation rates $J$ of $n$-decane as functions of supersaturation $S$ for various temperatures in logarithmic scale. Experimental data are taken from Rudek et\,al.\cite{Rudek1996} Markers and lines are defined in the same way as in Fig. \ref{fig1}.}
\label{fig4}
\end{figure}

We performed DGT and CNT computations for four alkanes; namely $n$-heptane, $n$-octane, $n$-nonane, and $n$-decane, for temperatures corresponding to nucleation rate experimental data by Rudek et\,al.,\cite{Rudek1996} Hung et\,al.,\cite{Hung1989} Luijten,\cite{Luijten1998} Viisaanen et\,al.\cite{Viisanen1998}, and Wagner and Strey,\cite{Wagner1984}
\begin{itemize}
 \item $n$-heptane (C7): 249 K, 259 K, 268 K, 276 K,
 \item $n$-octane (C8): 241 K, 248 K, 258 K, 267 K, 287~K, 298 K, 302 K,
 \item $n$-nonane (C9): 220 K, 230 K, 240 K, 247.6 K, 257 K, 267 K, 272 K, 284 K, 298 K, 313~K,
 \item $n$-decane (C10): 268 K, 276 K, 281 K, 294 K, 303~K, 316 K, 330 K.
\end{itemize}

A comprehensive experimental study of nucleation in $n$-alkanes from $n$-heptane to $n$-decane in nitrogen or argon has recently been provided by Ghosh et al.\cite{Ghosh_2010} We did not include it in the present comparisons because the very high nucleation rates (of the order of $10^{22}$ to $10^{23}$ $\mathrm{m^{-3} s^{-1}}$) make the comparison with the other data difficult. However, we note that their comparison with CNT prediction leads to similar conclusions, particularly concerning the strong temperature dependence of the deviation (cf. their Fig.~9).

Computations were done using PC-SAFT EoS and PR EoS. Influence parameters were computed using Eq. \eqref{eq_sigma_inf} based on the ``bare'' surface tension of the planar phase interface that was obtained from the experimental surface tension using Eq. \eqref{eq_sigma_bare}. (See Appendix \ref{app_sigma} for more information about the experimental surface tensions.) The computations of the density profiles for critical clusters were performed using the algorithm described in Sec. \ref{sec_method}.
The nucleation rates were evaluated using Eq. \eqref{eq_J_ic}, DGT works of formation were obtained using Eq. \eqref{eq_Delta_Omega1} for the computed density profiles, the works of formation for CNT were found using Eq. \eqref{eq_Delta_Omega2} with $\sigma=\sigma_\infty$.


Figs. \ref{fig1} to \ref{fig4} show nucleation rates depending on the supersaturations, Eq. \eqref{eq:S}, computed using the various combinations of DGT and CNT with PC-SAFT and PR EoSs compared with the experimental data in all figures. One color and one symbol always correspond to one temperature. Lines depict the theoretical data: solid lines with light-colored symbols are the DGT and PC-SAFT computations, dashed lines with light-colored symbols correspond to CNT and PC-SAFT, solid lines with white symbols to DGT and PR, dashed lines with white symbols to CNT and PR.

The darkest symbols depict the experimental data. The experiments were not conducted at the same nominal temperatures $T_0$ which correspond to the temperatures in the figures as the experimental temperatures $T_\r{exp}$ slightly varied. To enable a direct comparison, the experimental supersaturations $S_\r{exp}$
were corrected to match the nominal temperature.
A Taylor expansion was done for $\ln S_\r{exp}$ around the experimental temperature $T_\r{exp}$ to match the desired temperature $T_0$,
\begin{equation}
  \ln S_0 \,\dot{=}\, \ln S_{\mathrm{exp}} + \frac{\partial \ln S}{\partial T}(T_{\mathrm{exp}},J_{\mathrm{exp}})\cdot\big(T_0-T_{\mathrm{exp}}\big),
\end{equation}
where we defined $S_0=S (T_0, J_{\mathrm{exp}})$. However, the derivative ${\partial \ln S}/{\partial T}$ would be difficult to obtain, hence it was substituted by
\begin{equation}
 \frac{\partial \ln S}{\partial T} = -\,\frac{\partial \ln J}{\partial T}
\biggm/ \frac{\partial \ln J}{\partial \ln S}.
\end{equation}
These derivatives were evaluated by differentiating Eq. \eqref{eq_J_ic}.


The slopes of experimental data with respect to supersaturation relatively well correspond to the theoretical slopes. However, all the lines show a systematic temperature-dependent shift.

\begin{figure}
 \centering
 \includegraphics[scale=1]{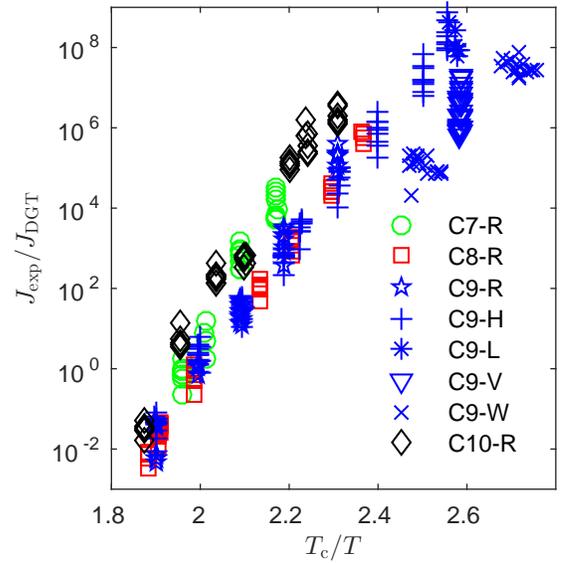}
\caption{Ratios of nucleation rates of the selected $n$-alkanes computed using DGT and PC-SAFT EoS and experimental nucleation rates as functions of inverse reduced temperature. Experimental data are taken from Rudek et al.\cite{Rudek1996} (R), Hung et\,al.\cite{Hung1989} (H), Luijten\cite{Luijten1998} (L), Viisaanen et\,al.\cite{Viisanen1998} (V) and Wagner and Strey \cite{Wagner1984} (W).}
\label{fig_ratio1}
\end{figure}

\begin{figure}
 \centering
 \includegraphics[scale=1]{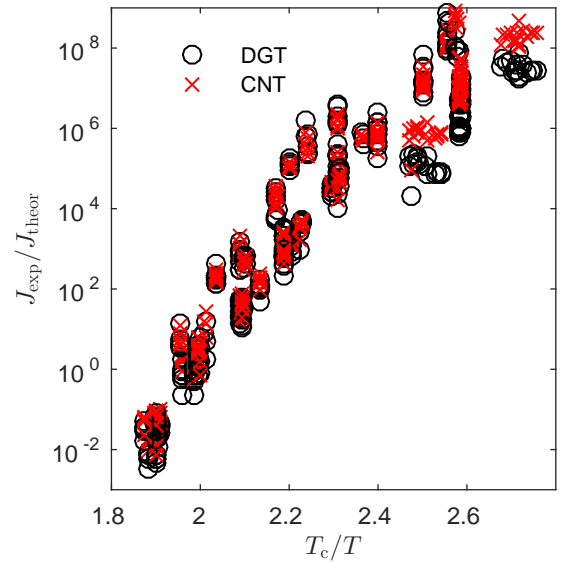}
\caption{Ratios of nucleation rates of the selected $n$-alkanes as functions of the inverse reduced temperature, comparison of results of DGT, shown in Fig. \ref{fig_ratio1}, and CNT, both combined with PC-SAFT EoS.}
\label{fig_ratio3}
\end{figure}

\begin{figure}
 \centering
 \includegraphics[scale=1]{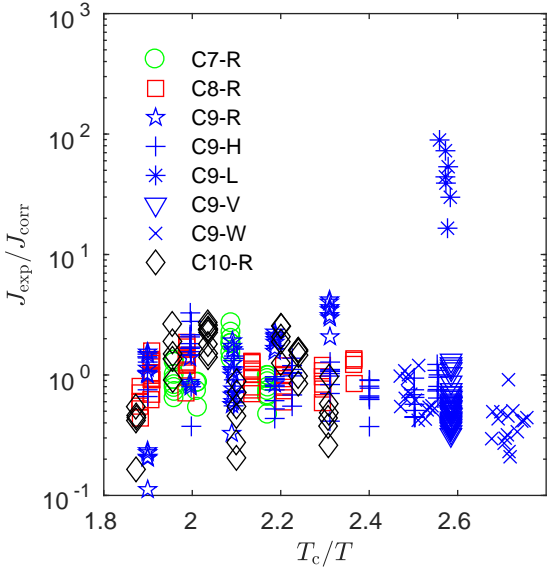}
\caption{Ratios of the experimental nucleation rates to DGT with PC-SAFT EoS predictions corrected using correlation, Eq. \eqref{eq_J_corr1}, as a function of the inverse reduced temperature.}
\label{fig_ratio4}
\end{figure}

\begin{figure}
 \centering
 \includegraphics[scale=1]{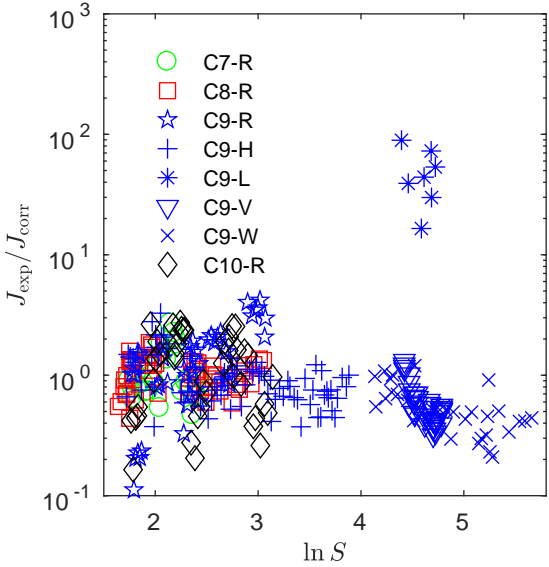}
\caption{Ratios of the experimental nucleation rates to DGT with PC-SAFT EoS predictions corrected using correlation, Eq. \eqref{eq_J_corr1}, as a function of supersaturation.}
\label{fig_ratio5}
\end{figure}

Another way how to study the quality of theoretical predictions is via ratios of experimental to predicted nucleation rates.
Figs. \ref{fig_ratio1} and \ref{fig_ratio3} show ratios of theoretical and experimental nucleation rates corresponding to the experimental temperatures and supersaturations as functions of inversed reduced temperatures.
Numerical data were twice interpolated using cubic splines; first, $\ln J$ was interpolated with respect to $\ln S$ on two theoretical isotherms neighboring $T_\r{exp}$ such that their $\ln S$ corresponded to the experimental value $\ln S_\r{exp}$. Then these two nucleation rates were again interpolated with respect to temperature to match $T_\r{exp}$.


Fig. \ref{fig_ratio1} gives the ratios of the experimental nucleation rates to the nucleation rates computed using DGT and PC-SAFT EoS. In the figure, one color and one symbol correspond to one experimental data set.
Results of DGT with PC-SAFT EoS and CNT with PC-SAFT EoS are compared in Fig. \ref{fig_ratio3}. The DGT results show slightly smaller temperature trend of the deviation from experimental data.

Clearly, the experimental data show a strong temperature trend which is not predicted by any of the theories. The ratios of experimental to theoretical nucleation rates for the various alkanes appear to follow almost a single linear function of the reciprocal reduced temperature,
\begin{equation}\label{ABeq}
\ln \left( \frac{J_\r{exp}}{J_\r{theor}}\right) \simeq A + B\frac{T_\r{c}}{T},
\end{equation}
where $A$ and $B$ are empirical parameters. Empirical correction, Eq.~(\ref{ABeq}), is of the same form as proposed by W\"{o}lk et al.\cite{Wolk2002} for correcting the CNT predictions of nucleation rates for water. As discussed by McGraw and Laaksonen\cite{McGraw1996}, the dependence of the ratio of experimental to theoretical nucleation rates on supersaturation is rather weak; in other words, the dependence of the nucleation rate on the supersaturation is predicted relatively well by the theory.
However, some discrepancies can be observed for the data of $n$-nonane, for which the experiments were done in a wide range of supersaturations. A failure of CNT to properly describe the supersaturation dependence accurately has been discussed by Girshick.\cite{Girshick_2014} We propose an additional term describing the dependence on the supersaturation,
\begin{equation}
\ln \left( \frac{J_\r{exp}}{J_\r{theor}}\right) \simeq A + B\frac{T_\r{c}}{T} + C \ln S.
\label{eq_J_corr1}
\end{equation}
We used the least-squares method to determine the parameters in Eq. \eqref{eq_J_corr1} for $n$-nonane for both nucleation theories and both EoSs. Except for $n$-nonane, the nucleation rate data for selected $n$-alkanes are available in a rather narrow supersaturation range. This makes it impossible to obtain a reliable value for parameter $C$. Therefore, we used the numerical value of coefficient $C$ for $n$-nonane as an approximation of the supersaturation dependence for other alkanes.
Figs. \ref{fig_ratio4} and \ref{fig_ratio5} show the ratios of experimental nucleation rates $J_\r{exp}$ to predictions of Eq. \eqref{eq_J_corr1}. These figures show that the correlation successfully captures the variability of the experimental data. Values of coefficients $A, B$, and $C$ are listed in Tab. \ref{tab_corrs} for DGT and CNT combined with PC-SAFT and PR EoSs.

\begin{table}[ht]
\caption{Coefficients of Eq. \eqref{eq_J_corr1}.}
\label{tab_corrs}
\centering
 \begin{tabular}{|c|ccc|}
\hline
DGT with PC-SAFT: &       A   &     B   &    C    \\
\hline
$n$-heptane & -110.7738 & 62.1559 & -6.1818 \\
 $n$-octane & -94.01211 & 53.1095 & -6.1818 \\
 $n$-nonane & -91.89308 & 52.5203 & -6.1818 \\
 $n$-decane & -100.6096 & 58.3962 & -6.1818 \\
\hline
CNT with PC-SAFT: &       A   &     B   &    C    \\
\hline
$n$-heptane & -97.9379 & 53.8308 & -3.7779 \\
 $n$-octane & -82.0400 & 45.0167 & -3.7779 \\
 $n$-nonane & -79.5862 & 44.0546 & -3.7779 \\
 $n$-decane & -88.9715 & 50.0613 & -3.7779 \\
\hline
DGT with PR: &       A   &     B   &    C    \\
\hline
$n$-heptane & -134.7050 & 80.0453 & -12.0051 \\
 $n$-octane & -119.4497 & 73.1246 & -12.0051 \\
 $n$-nonane & -119.5625 & 75.3177 & -12.0051 \\
 $n$-decane & -130.1146 & 83.2105 & -12.0051 \\
\hline
CNT with PR: &       A   &     B   &    C    \\
\hline
$n$-heptane & -124.5077 & 72.4331 & -8.7391 \\
 $n$-octane & -109.2277 & 65.3503 & -8.7391 \\
 $n$-nonane & -109.5674 & 67.5294 & -8.7391 \\
 $n$-decane & -119.1090 & 74.7644 & -8.7391 \\
\hline
 \end{tabular}
\end{table}

\begin{figure}
 \centering
 \includegraphics[scale=1]{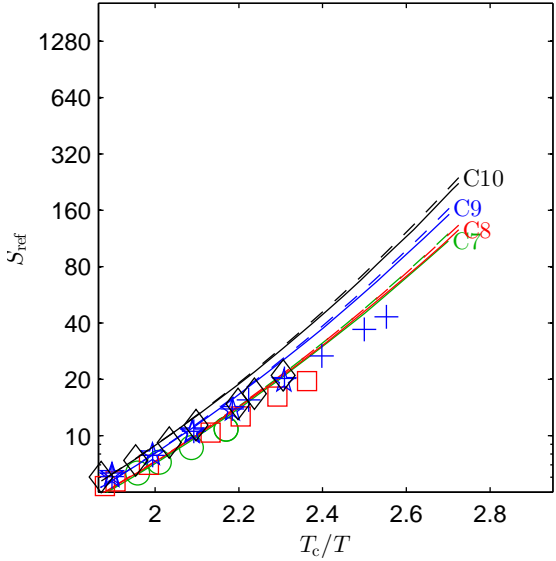}
\caption{Dependence of supersaturation $S$ on the inverse reduced temperature $T_\r c/T$ for four selected $n$-alkanes fitted to a constant nucleation rate of $J_\r{ref} = 7.57 \times 10^4 \r m^{-3}\r s^{-1}$. Lines correspond to the DGT predictions: solid lines to PC-SAFT EoS, dashed lines to PR EoS. Symbols depict the experimental data\cite{Hung1989,Rudek1996} (same correspondence as in Fig. \ref{fig_ratio1}). One color corresponds to one substance.}
\label{fig_S1}
\end{figure}
\begin{figure}
 \centering
 \includegraphics[scale=1]{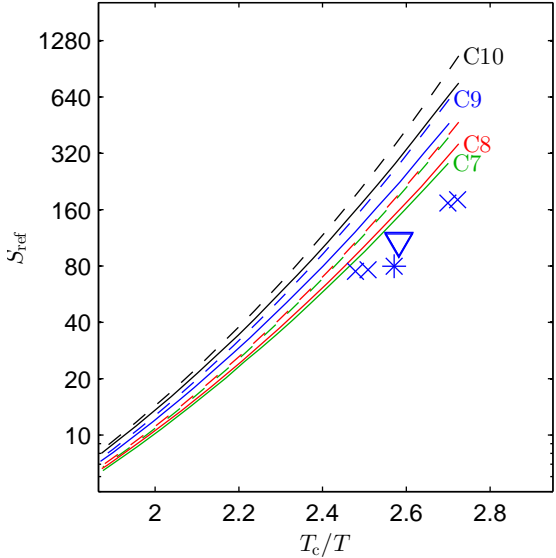}
\caption{Dependence of supersaturation $S$ on the inverse reduced temperature $T_\r c/T$ for four selected $n$-alkanes fitted to a constant nucleation rate of $J_\r{ref} = 3.22 \times 10^{13} \r m^{-3}\r s^{-1}$. Lines are defined in the same way as in Fig. \ref{fig_S1}. Symbols depict the experimental data\cite{Luijten1998,Viisanen1998,Wagner1984} (same correspondence as in Fig. \ref{fig_ratio1}).}
\label{fig_S2}
\end{figure}
\begin{figure}
 \centering
 \includegraphics[scale=1]{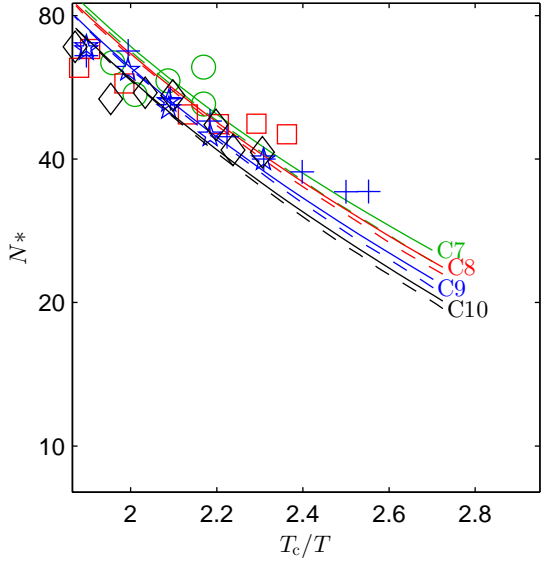}
\caption{Dependence of number of molecules in critical cluster $N^*$ on the inverse reduced temperature $T_\r c/T$ for four selected $n$-alkanes fitted to a constant nucleation rate of $J_\r{ref} = 7.57 \times 10^4 \r m^{-3}\r s^{-1}$. Lines and experimental data are defined in the same way as in Fig. \ref{fig_S1}.}
\label{fig_N1}
\end{figure}
\begin{figure}
 \centering
 \includegraphics[scale=1]{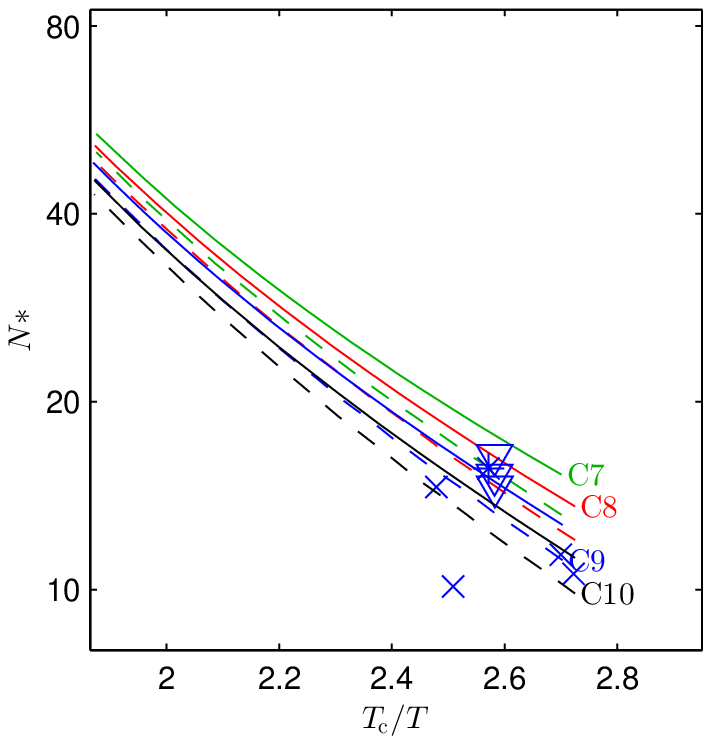}
\caption{Dependence of number of molecules in critical cluster $N^*$ on the inverse reduced temperature $T_\r c/T$ for four selected $n$-alkanes fitted to a constant nucleation rate of $J_\r{ref} = 3.22 \times 10^{13} \r m^{-3}\r s^{-1}$. Lines are defined in the same way as in Fig. \ref{fig_S1} and experimental data as in Fig. \ref{fig_S2}.}
\label{fig_N2}
\end{figure}

Typically, the experiments provide nucleation rates in a relatively narrow range determined by the experimental method. Therefore, it is practical to represent experimental data as a temperature dependence of supersaturation $S_\r{ref}$ corresponding to a chosen reference nucleation rate $J_\r{ref}$.
Figs. \ref{fig_S1} and \ref{fig_S2} show a temperature dependence of the supersaturation depending on the inverse reduced temperature $T_\r c/T$ for the selected four substances. Numerical calculations were performed in a wider range of temperatures compared to experimental data in order to provide a better picture of the temperature trend. Experimental data by Rudek et al.\cite{Rudek1996} and Hung et al.\cite{Hung1989} were measured in a substantially lower range of nucleation rates than by Luijten,\cite{Luijten1998} Viisaanen et\,al.,\cite{Viisanen1998} and Wagner and Strey \cite{Wagner1984}. For this reason, we used two reference nucleation rates $J_\r{ref} = 7.57 \times 10^4 \r m^{-3}\r s^{-1}$ and $J_\r{ref} = 3.22 \times 10^{13} \r m^{-3}\r s^{-1}$, which were obtained as geometrical averages of the nucleation rates included in each data subset. We note that Figs. \ref{fig_S1} and \ref{fig_S2} have have the same ranges of both $x$ and $y$ axes. 

In order to reduce the experimental data obtained for various nucleation rates to a given reference nucleation rate $J_\r{ref}$, the theoretical data were interpolated using a cubic spline. For experimental data, we assumed a linear dependence,
\begin{equation}
 \ln J_{\mathrm{exp}} = b_1 + b_2 \ln S_{\mathrm{exp}},
 \label{eq_lnJb1b2}
\end{equation}
where coefficients $b_1$ and $b_2$ were fitted using the least squares method for each isothermal experimental data set. One symbol in Figs. \ref{fig_S1} and \ref{fig_S2} (and later also in Figs. \ref{fig_N1} and \ref{fig_N2}) corresponds to one isothermal data set of the measurements, typically including five measurements. Supersaturations $S$ range from 5.39 to 43.18 for the case of $J_\r{ref} = 7.57 \times 10^4 \r m^{-3}\r s^{-1}$ (Fig. \ref{fig_S1}) and from 74.94 to 181.525 for the case with $J_\r{ref} = 3.22 \times 10^{13} \r m^{-3}\r s^{-1}$ (Fig. \ref{fig_S2}).

Figs. \ref{fig_N1} and \ref{fig_N2} show a temperature dependence of the number of molecules in the critical cluster $N^*$ depending on the inverse reduced temperature $T_\r c/T$. The data was fitted to match one reference nucleation rate as discussed above. For the DGT predictions, the number of molecules $N^*$ was evaluated as
\begin{equation}
  N^* = \rho_\mathrm{V}V_\mathrm{N^*} + \Delta N_\mathrm{N^*},
\end{equation}
where $\Delta N_\mathrm{N^*}$ is the excess number of molecules given by Eq. \eqref{eq_deltaN} and $V_\mathrm{N^*}$ is the volume of the droplet consisting of $N^*$ molecules. A rational estimate of volume is the volume of a sphere enclosed by the surface of tension, $V_\mathrm{N^*} = 4/3 \pi r_\mathrm{s,N^*}^3$. 

Regarding the experimental data, the number of molecules of the critical cluster was evaluated using the nucleation theorem,\cite{Oxtoby1994}
\begin{equation}
  \left(\frac{\partial \ln J}{\partial \ln S}\right)_\r T = N^* + 1.
\end{equation}
As the derivative can be fitted by Eq. \eqref{eq_lnJb1b2}, $N^* = b_2-1$.

Again, we split the experimental data into two groups and used reference nucleation rates $J_\r{ref} = 7.57 \times 10^4 \r m^{-3}\r s^{-1}$ and $J_\r{ref} = 3.22 \times 10^{13} \r m^{-3}\r s^{-1}$. The number of molecules in critical cluster $N^*$ in Figs. \ref{fig_N1} and \ref{fig_N2} range from 34.2 to 68.9 for the case of $J_\r{ref} = 7.57 \times 10^4 \r m^{-3}\r s^{-1}$ and from 10.1 to 16.4 for the case with $J_\r{ref} = 3.22 \times 10^{13} \r m^{-3}\r s^{-1}$.

As can be seen in Figs. \ref{fig_S1}, \ref{fig_S2}, \ref{fig_N1}, and \ref{fig_N2}, the theories provide qualitatively correct predictions, despite the large disagreement in terms of the nucleation rate. However, a noticeable disagreement can be seen in the slopes. Predicted derivatives of the supersaturation $S$ and of the number of molecules in the critical cluster $N^*$ with respect to temperature at a constant nucleation rate differ from the experimental slopes, especially at low temperatures.

\begin{figure}
 \centering
 \includegraphics[scale=1]{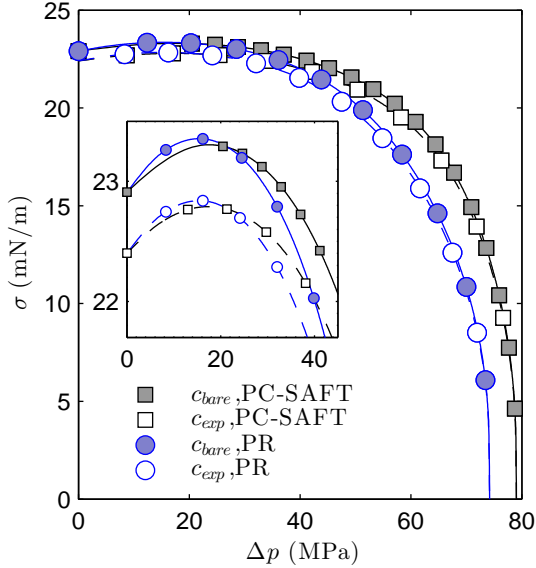}
\caption{Surface tension $\sigma$ as a function of the Laplace pressure $\Delta p$ of $n$-heptane at temperature $T = 276 \,\r K$. Lines are results of DGT combined with PR and PC-SAFT EoSs using the influence parameter $c$ obtained from experimental surface tension $\sigma_{exp}$, Eq. \eqref{eq_sigma_inf}, and bare surface tension $\sigma_{bare}$, Eqs. \eqref{eq_sigma_inf} and \eqref{eq_sigma_bare}. Inset is the detail of the beginning where the lines cross.}
\label{fig_sigma_cw0_cw1}
\end{figure}

\begin{figure}
 \centering
 \includegraphics[scale=1]{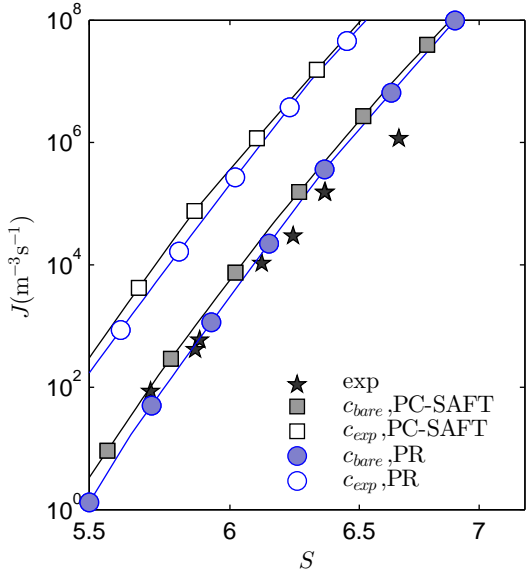}
\caption{Nucleation rates $J$ as functions of supersaturation $S$ of $n$-heptane at temperature $T = 276 \,\r K$. Lines are the same as in Fig. \ref{fig_sigma_cw0_cw1}. Nucleation rates are also compared to the experimental data by Rudek et al.\cite{Rudek1996} (black stars).}
\label{fig_J_cw0_cw1}
\end{figure}

As discussed in Sec. V, CW broaden the phase interface and somewhat decrease the experimental surface tension with respect to the bare surface tension corresponding to a phase interface constrained to be free of CW.
Fig. \ref{fig_sigma_cw0_cw1} shows the surface tension depending on the Laplace pressure $\Delta p$ computed using DGT combined with the two EoSs (PR and PC-SAFT), and with two different definitions of the influence parameters. The figure demonstrates an example for $n$-heptane at temperature of $276$~K.
The Laplace pressure is connected with the surface-of-tension radius of the droplet $r_s$ via the Young--Laplace equation, Eq. \eqref{eq_young}. Therefore, $\Delta p = 0$ Pa at left corresponds to an infinite radius, i.e., the planar phase interface with surface tension $\sigma_\infty$. The size-dependent surface tension was obtained from Eq. \eqref{eq:DOcnt} based on the work of formation, Eq. \eqref{eq_Delta_Omega1}. At the right end, the surface tension according to DGT vanishes at the the spinodal.
Lines $c_\r{bare}$ connect results for influence parameters computed using Eq. \eqref{eq_sigma_inf}, where $\sigma_\infty$ was cleared of CW using Eq. \eqref{eq_sigma_bare}. This is a prediction considered as correct, because, as discussed in Sec. V, the surface of the critical clusters is practically free of CW and, at the same time, the bare surface tension is the appropriate value to compute the influence parameter as DGT does not include CW. Lines $c_\r{exp}$ correspond to the influence parameters computed via Eq. \eqref{eq_sigma_inf} using the experimental surface tension. This is the standard approach; however, it is considered as less accurate.
From Fig. \ref{fig_sigma_cw0_cw1} it can be seen that $\sigma_\r{bare} > \sigma_\r{exp}$. The surface tension curves for both EoSs start at either $\sigma_\r{bare}$ or at $\sigma_\r{exp}$ at $\Delta p = 0$ Pa, but then tends to a spinodal value given by the particular EoS. The inset figure shows a detail of the trends for low Laplace pressures. The range of Laplace pressures for the critical clusters in the experiments is on the descending branches of the curves close to the maximum. In the case shown in Fig. \ref{fig_sigma_cw0_cw1}, the Laplace pressures range from 27.3 to 29.9~MPa, corresponding to critical radii of 1.64 to 1.50~nm. It can be seen that CW have a significant effect on the surface tension. This effect is bigger or at least comparable (in other cases) with the effect of replacing the cubic PR EoS by the more accurate PC-SAFT EoS. A reason can be seen in a crossing of the lines predicted by both EoSs just in the relevant range.

Fig. \ref{fig_J_cw0_cw1} shows nucleation rates depending on the supersaturation of $n$-heptane at temperature of $276$~K. Again, the two EoSs and two different values for the influence parameters were used, similarly as in Fig. \ref{fig_sigma_cw0_cw1}. The stars depict the experimental data. Constraining the phase interface to be free of CW increases the surface tension and, consequently, enhances the work of formation and lowers the nucleation rates. The effect is quite large, which proves the importance of such a treatment. We note that the quantitative agreement of the experimental data with the theoretical prediction shown in Fig. \ref{fig_J_cw0_cw1} is rather fortuitous: due to the strong temperature effect, the differences are larger at other temperatures.

\section{Conclusions}
\label{sec_conclusions}
We computed nucleation rates of four alkanes: $n$-heptane, $n$-octane, $n$-nonane, and $n$-decane, using the density gradient theory complemented by an original treatment of capillary waves, and the classical nucleation theory. For both approaches we used a simple cubic Peng--Robinson EoS, which essentially is an empirical modification of the van der Waals EoS, and the PC-SAFT EoS, which is a modern EoS based on molecular arguments. The computed predictions were compared to the experimental nucleation rate data.

An original algorithm was developed to solve an Euler--Lagrange equation of DGT which together with the boundary conditions forms a boundary value problem. Several enhancements of the numerical method were developed in order to ensure a robust convergence to the physically-proper density profiles.

For computing the nucleation rates, we used an expression based on the classical nucleation kinetics with corrected supersaturation dependence, Eq. \eqref{eq_J_ic}. The work of formation was obtained using DGT combined with our CW treatment or, in case of CNT, using the Gibbs formula based on the surface tension for the planar phase interface.

In our investigation of the effect of capillary waves on the surface tension and nucleation rates we found that the critical clusters are so small that the shortest wavelengths of CW are already too large to fit to the surface of the nanoscopic droplet representing the critical cluster. This means that the thermal motion of the molecules and the corresponding fluctuations of density and other local quantities are completely accounted for in the DGT description. Therefore, for a consistent computation of the works of formation using either DGT or CNT, it is appropriate to remove the effect of CW from the planar surface tension used in the computations. This was achieved using Eq.~(\ref{eq_sigma_cw1}) based on the theory of Meunier.\cite{Meunier1987}  In DGT, the influence parameter was fitted to reproduce the bare surface tension corresponding to the planar phase interface free of CW.

Three contributions to the surface tension of a droplet can be recognized in Fig. \ref{fig_sigma_cw0_cw1}. First is the Tolman's linear effect which accounts for the mild increase of surface tension (for low Laplace pressure $\Delta p$). This effect is related to symmetry properties of the density profile, expressed by the Tolman length. Second, quadratic effect, appears to be decisive in bending the curvature such that the surface tension reaches a maximum and then decreases towards the spinodal. The quadratic effect can be attributed to the finite thickness of the phase interface and to changes in its shape as the critical cluster becomes smaller. Third is the CW effect which decreases the surface tension of nano-droplets as a part of the spectrum of the CW that do not fit to the finite dimensions of the droplet. Suppression of CW leads to a considerable enhancement of the predicted surface tensions and, in turn, to a decrease in the nucleation rates.

A pre-requisite for realistic modeling using DGT is an accurate EoS. The PC-SAFT EoS provides a more accurate representation of thermodynamic properties in the stable liquid and gas regions covered with experimental data. In the Supplementary Material 2, we provide some comparisons showing that PC-SAFT EoS is clearly superior to the PR EoS, which is generally considered as a reasonable standard for modeling hydrocarbon systems. Besides, the molecular foundations of PC-SAFT EoS allow to consider the predictions for the metastable and unstable regions as plausible. Fig.~2 in the Supplementary Material 2 shows that the van der Waals loops predicted by both EoSs differ substantially. Despite that, the difference in the predicted surface tensions of critical clusters and, consequently, in the predicted nucleation rates, is not tremendous, at least for the systems studied here. The reason is shown in the inset of Fig.~16: Incidentally, the curves predicted by DGT combined with PC-SAFT EoS intersect in the region corresponding to the critical clusters.

This study has been performed with the goal of possibly finding a reason for a disagreement of CNT with experimental data, which is particularly striking in the case of $n$-alkanes. In Figs.~3 to 6, we provide comparison of theoretical predictions with experimental nucleation rate data. The differences between the DGT/CW and CNT predictions is significant, reaching several orders of magnitude in the nucleation rate. Also the difference between predictions by the different EoSs is appreciable. Despite that, the huge deviation from the experimental data, particularly in the temperature dependence, was not remedied. To characterize the deviation from experiments, various approaches can be adopted. One can deduce the magnitude of the microscopic surface tension (corresponding to the critical cluster) from nucleation rate data. For example, Baidakov and Skripov\cite{BS_1982} found that for bubble nucleation in several liquified gases it is appropriate to reduce the macroscopic surface tension by a few percent. In the present case, this approach does not appear to be insightful. Due to the strong temperature dependence, the correction to the surface tension would change from large positive values at low temperature to large negative corrections at the high-temperature limit of the experimental window. In section~\ref{sec_nucleation} we noted that there are several modifications to CNT differing in the pre-exponential factor. The quantity $\vartheta$, the ``translational volume scale'',\cite{RKK_1998} remains, despite numerous theoretical contributions,\cite{Reguera2004} a rather uncertain quantity. For this reason, we consider as more appropriate to represent the deviation in terms of a multiplicative correction to the predicted nucleation rate. We provide this correction in form of Eq.~(\ref{eq_J_corr1}) for all combinations of theories (DGT/CW, CNT) and equations of state (PC-SAFT, PR) to facilitate further theoretical analyses as well as practical computations. The dominant term is the temperature correction. We also included a correction to the supersaturation dependency which is weaker and difficult to asses, because it relies on comparison of various experimental methods, which might be subject to systematic offsets.

We presented partial improvements to the classical nucleation theory which bring the modeling closer to reality. However, the resolution of the temperature-dependent deviation,\cite{Malila2008} which is very strong in case of $n$-alkanes, remains unresolved.

\section*{Supplementary material}
The data points of the figures presented in this article are listed in the Supplementary Material 1. Additional figures are included in Supplementary Material 2 that compare saturated properties of the four $n$-alkanes for both EoSs and justify superiority of PC-SAFT EoS over PR EoS. Further figures of nucleation rates are also included.

\begin{acknowledgments}
The research leading to these results has received funding from the Norwegian Financial
Mechanism 2009-2014 under Project Contract No. 7F14466, from the Czech Science Foundation
grant No. GJ15-07129Y, and from the institutional support RVO: 61388998. Besides, B.P. thanks
for a partial support from the grant No. SGS15/215/OHK4/3T/14.
\end{acknowledgments}

\appendix
\section{Parameters of equations of state}
For PC-SAFT EoS, molecular parameters $m, \sigma$, and $\epsilon$ for selected $n$-alkanes used in this study were taken from the original work of Gross and Sadowski\cite{Gross2001}.

The Peng--Robinson equation is given by
\begin{equation}
 p = \frac{\rho R_\r g T}{1-b\rho} - \frac{a\alpha\rho^2}{1+2b\rho-(b\rho)^2},
\end{equation}
where $R_\r g = 8.3144598 \,\r{J\, mol^{-1}\, K^{-1}}$ is the universal gas constant and $a, b, \alpha$ are Peng--Robinson constants given as follows
\begin{eqnarray}
 a &=& 0.45724 \frac{R_\r g^2 T_\r c^2}{p_\r c}, \nonumber\\
 b &=& 0.0778 \frac{R_\r g T_\r c}{p_\r c}, \\
 \alpha &=& \Big[ 1 + \big(1-\sqrt{T/T_\r c}\big) (0.37464 + 1.54226\omega-0.26992\omega^2) \Big]^2\nonumber,
\end{eqnarray}
where $p_\r c$ is the critical pressure and $\omega$ is the acentric factor. Values fot he acentric factor were taken from the work by Lemmon and Goodwin.\cite{Lemmon2000} Critical parameters $T_\r c$ and $p_\r c$ and acentric factor $\omega$ for four selected $n$-alkanes are listed in Tab. \ref{tab_pr_pars}.



\begin{table}
  \caption{Parameters of PR EoS}
  \label{tab_pr_pars}
  \begin{tabular}{ |c|ccc| }
    \hline
             & $T_\r c \, (\r K)$ & $p_\r c \, (\r{bar})$ & $\omega (-)$ \\
    \hline
     $n$-heptane  & 540.13 & 27.36 & 0.3365 \\
      $n$-octane  & 569.32 & 24.97 & 0.3765 \\
      $n$-nonane  & 594.55 & 22.81 & 0.4140 \\
       $n$-decane & 617.70 & 21.03 & 0.4495 \\
    \hline
  \end{tabular}
\end{table}

\section{Experimental surface tension}
\label{app_sigma}
The influence parameter $c$ was computed using the experimental surface tension. The experimental values were taken from Jasper et al.\cite{Jasper1953} for all the selected substances, Grigoryev\cite{Grigoryev1992} for $n$-heptane (C7) and $n$-octane (C8), Voliak and Andreeva\cite{Voliak1961} for C7 and C8, Rolo et al.\cite{Rolo2002} for C7 and $n$-decane (C10), Okada et al.\cite{Okada1990} for C8.

The actual value of the surface tension was computed using a temperature correlation. For $n$-heptane and $n$-octane, we used approach by Somayajulu\cite{Somayajulu1988},
\begin{equation}
 \sigma = A \bigg(\frac{T_\r c}{T}-1\bigg)^{5/4} + B \bigg(\frac{T_\r c}{T}-1\bigg)^{9/4} + C \bigg(\frac{T_\r c}{T}-1\bigg)^{13/4}.
\label{eq_exp_sigma1}
\end{equation}
For $n$-nonane and $n$-decane we developed our own fit based on the scaling around the critical temperature,
\begin{equation}
 \sigma = A \bigg(\frac{T_\r c}{T}-1\bigg)^{1.26}.
\label{eq_exp_sigma2}
\end{equation}
Values of parameters $A$, $B$, and $C$ from Eqs. \eqref{eq_exp_sigma1} and \eqref{eq_exp_sigma2} are listed in Tab. \ref{tab_sigma}. 
\begin{table}
   \caption{Table of parameters for the fit of the experimental surface tension}
   \label{tab_sigma}
   \begin{tabular}{ |c|ccc| }
     \hline
             & $A$ (mN/m) & $B$ (mN/m) & $C$ (mN/m) \\
    \hline
     $n$-heptane  & 54.1778 & -0.7586 & 3.9897 \\
      $n$-octane  & 56.5399 & -10.4928 & 8.4723 \\
      $n$-nonane  & 53.9000 & - & - \\
       $n$-decane & 53.6000 & - & - \\
    \hline
   \end{tabular}
\end{table}

\bibliographystyle{aapmrev4-1}
\bibliography{reference}

\end{document}